\documentclass{iaus}
\usepackage{graphicx}
\usepackage{natbib}
\usepackage{pagefram}
\title[Radio PNe in the MCs] 
{Radio Planetary Nebulae in \\
the Magellanic Clouds}

\author[E.J. Crawford et al.]   
{Evan~J.~Crawford.$^1$, Miroslav~D.~Filipovi\'c$^1$, Ivan~S.~Boji\v{c}i\'c$^{2,3,4}$, Martin~Cohen$^5$, Jeff~L.~Payne$^1$, Ain~Y. De~Horta$^1$, Warren~Reid$^{2,3}$}

\affiliation{$^1$University of Western Sydney, Locked Bag 1797, Penrith NSW 2751 Australia\\ email: {\tt e.crawford@uws.edu.au m.filipovic@uws.edu.au} \\[\affilskip]
$^2$Department of Physics and Astronomy, Macquarie University, Sydney, NSW 2109, Australia \\[\affilskip]
$^3$Macquarie University Research Centre in Astronomy, Astrophysics \& Astrophotonics\\[\affilskip]
$^4$Australian Astronomical Observatory, PO Box 296, Epping, NSW\\[\affilskip]
$^5$Radio Astronomy Laboratory, University of California, Berkeley, CA 94720}

\pubyear{2011}
\volume{283}  
\pagerange{1--2}
\jname{Planetary Nebulae an Eye to the Future}
\editors{A.C. Editor, B.D. Editor \& C.E. Editor, eds.}
\begin{document}

\maketitle

\begin{abstract}
We present preliminary results of our deep Australia Telescope Compact Array (ATCA) radio-continuum survey of the Magellanic Clouds Planetary Nebulae.

\keywords{galaxies: Magellanic Clouds --- radio-continuum: galaxies --- ISM: planetary nebulae}
\end{abstract}

\firstsection 
\section{Introduction}

The importance of the radio-continuum properties of planetary nebulae (PNe) has been recently reinstated with the report of the radio-continuum observations of PNe in the Magellanic Clouds \citep{2009MNRAS.399..769F, 2008SerAJ.176...65P, 2008SerAJ.177...53P}. They report the extragalactic radio-continuum detection of 15 PNe in both Clouds using Australia Telescope Compact Array and Parkes mosaic surveys \citep{1995A&AS..111..311F, 1997A&AS..121..321F, 2002MNRAS.335.1085F, 2007MNRAS.382..543H}. Four of the PNe are located in the Small Magellanic Cloud (SMC) and 11 are located in the Large Magellanic Cloud (LMC). \citet{2010SerAJ.181...63B} reported one additional radio detection of the SMC PNe -- SMP~S24. Prior to these studies, a radio detection of only three extragalactic PNe have been reported in the literature \citep{ZHCL94,DPZW2000}. Based on the radio-continuum properties of radio-bright Galactic PNe the expected radio flux densities at the distance of the LMC/SMC are up to $\sim$2.5 and $\sim$2.0~mJy at 1.4~GHz, respectively. The reported Magellanic Clouds (MCs) radio PNe detections represent only $\sim$3\% of the optical PNe population of the MCs. Most likely, we are selecting only the strongest radio-continuum emitters, possibly at a variety of different stages of their evolution \citep{2009A&A...503..855V}.

The known and well refined distances to the MCs provide a great opportunity for the accurate evaluation of important physical properties for PNe such as ionised mass and electron densities.  Also, a statistically significant sample of radio detected extragalactic PNe will allow the construction and examination of the bright end of the radio PN luminosity function (PNLF) and comparison with established theoretical and empirical MCs PNLF's obtained at other wavelengths \citep{1995ApJ...452..515S,2002AJ....123..269J}. Therefore, we initiated a deep, 6-cm radio-continuum survey which will attempt to detect and accurately measure the radio-continuum flux densities of $\sim50$ MCs PNe. Our positive radio detection of 39 PNe from MCs opens a door to the exciting possibility of construction of the PNe radio luminosity function which would be of extreme help in understanding of properties of PNe in our own galaxy. Our final aim is to collect radio data for all radio detectable MCs PNe. However, in this stage, it is important that we measure the accurate flux densities for a significant sample of MC PNe which can be later on properly correlated to the available multi-wavelength data, and consequently allow us to predict the best strategies for the following large scale observational studies of radio-continuum properties of the MCs PNe sample.

\begin{figure}
 \begin{center}
  \includegraphics[scale=0.29,angle=-90]{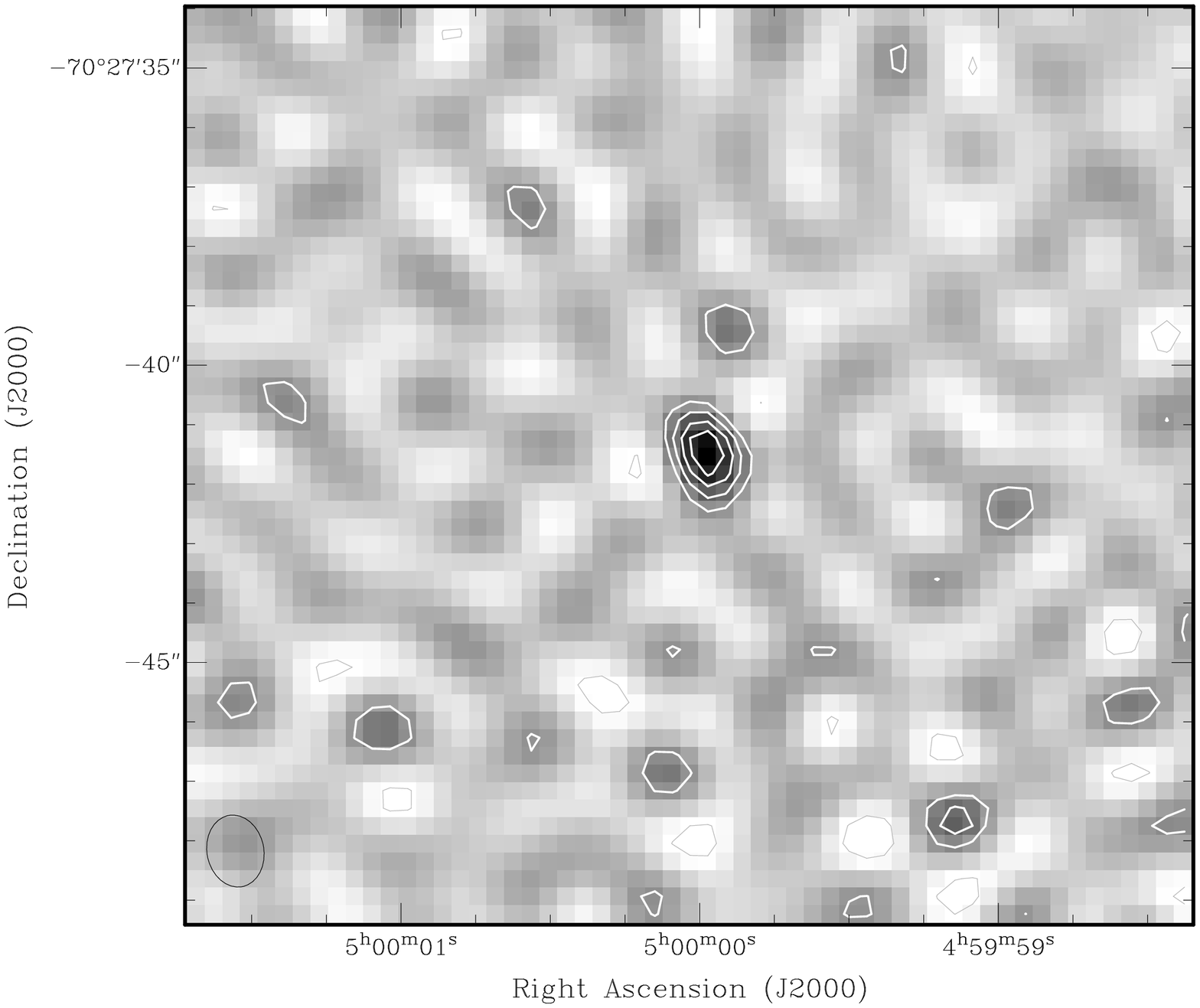}
  \includegraphics[scale=0.29,angle=-90]{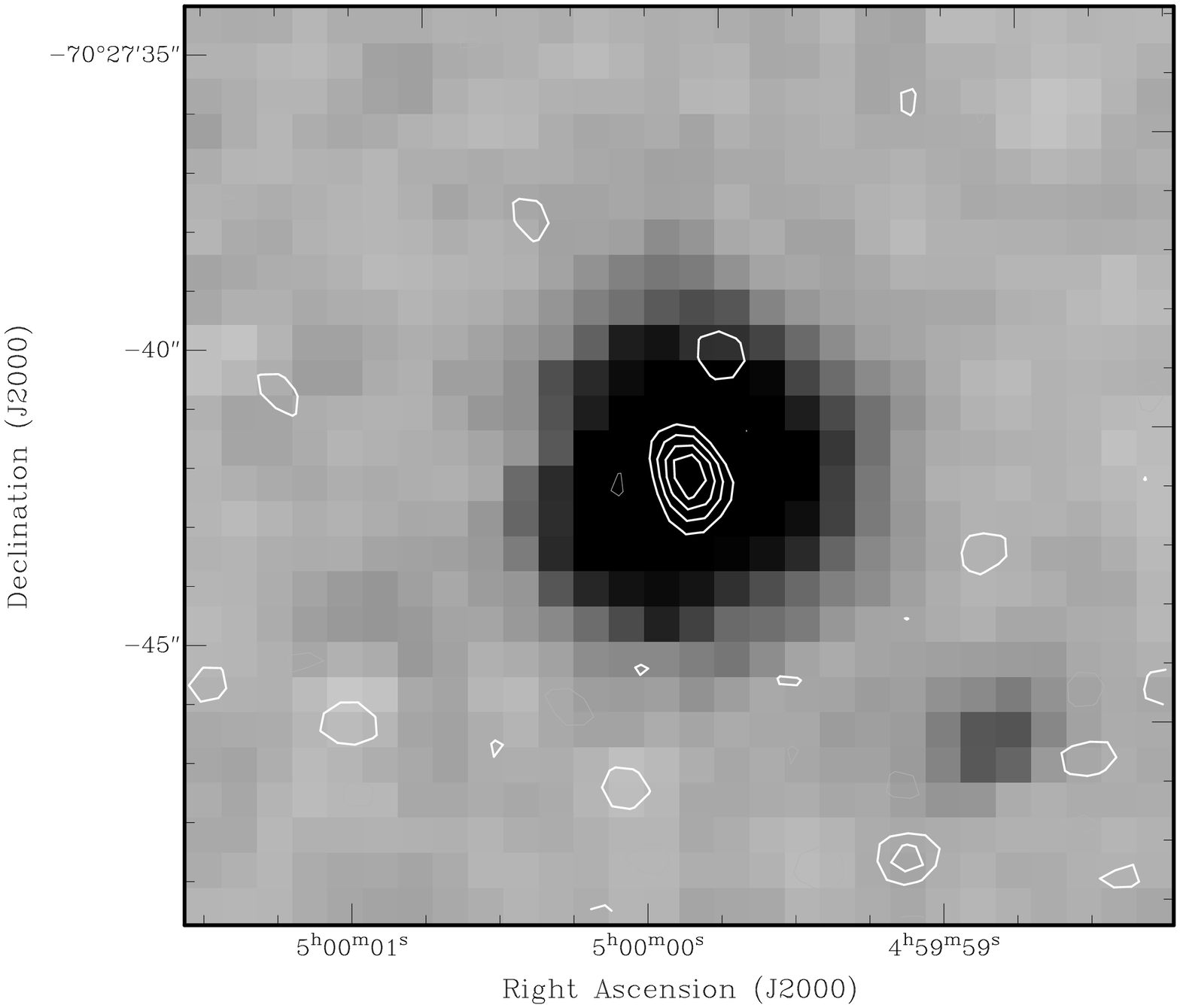}
  \caption{An example of the finding charts showing {\it left}: 6~cm total intensity radio image and {\it right}: 5.8$\mu m$ SAGE LMC image \citep{2006AJ....132.2268M} of one of the radio-faintest PN detected in our survey --- SMP~13. Images are overlaid with contours from for the radio continuum image at -0.15, 0.15, 0.25 and 0.35 mJy/beam.}
  \label{figs}
 \end{center}
\end{figure}

\section{Observations and preliminary results}

The new sample was observed with the the Australia Telescope Compact Array (ATCA), as part of project C2367, on 28$^\mathrm{th}$-- 30$^\mathrm{th}$ November 2010, with an array configuration 6C, at wavelengths of 3 and 6~cm ($\nu$=9000 and 5500~MHz). We made use of the recently upgraded receiver system (The Compact Array Broadband Backend; CABB) which significantly improves the observing capabilities of the ATCA. The observations were carried out in the so called ``snap-shot'' mode, totaling $\sim$1 hour of integration over a 12 hour period.  Our preliminary 6\,cm results show great improvements with some 23 new PNe clearly detected (above the 3$\sigma$ level; Fig.~\ref{figs}) bringing the total number of radio detected PNe in the MCs to 39 of which 34 are from the LMC.

\end{document}